\documentclass[preprint,superscriptaddress,aps]{revtex4}

\usepackage{graphicx}
\usepackage{dcolumn}
\usepackage{bm}

\begin{document}

\preprint{}

\title{Fine Structure of the $1s3p\;^3{\rm P}_{J}$ Level in Atomic
$^4$He:\\
Theory and Experiment}

\author{P. Mueller}
\email{pmueller@anl.gov} \affiliation{Physics Division, Argonne
National Laboratory, Argonne, Illinois 60439, USA}
\author{L.-B. Wang}
\affiliation{Physics Division, Argonne National Laboratory, Argonne,
Illinois 60439, USA} \affiliation{Physics Department, University of
Illinois at Urbana-Champaign, Urbana, Illinois 61801, USA}
\author{G. W. F. Drake}
\email{gdrake@uwindsor.ca} \affiliation{Physics Department, University
of Windsor, Windsor, Ontario, Canada N9B 3P4}
\author{K. Bailey}
\affiliation{Physics Division, Argonne National Laboratory, Argonne,
Illinois 60439, USA}
\author{Z.-T. Lu}
\affiliation{Physics Division, Argonne National Laboratory, Argonne,
Illinois 60439, USA}
\author{T. P. O'Connor}
\affiliation{Physics Division, Argonne National Laboratory, Argonne,
Illinois 60439, USA}

\date{July 21, 2004}

\begin{abstract}
We report on a theoretical calculation and a new experimental
determination of the $1s3p\;^3{\rm P}_{J}$ fine structure intervals in
atomic $^4$He. The values from the theoretical calculation of
8113.730(6)~MHz and 658.801(6)~MHz for the $\nu_{01}$ and $\nu_{12}$
intervals, respectively, disagree significantly with previous
experimental results. However, the new laser spectroscopic measurement
reported here yields values of 8113.714(28)~MHz and
658.810(18)~MHz for these intervals. These results show an excellent
agreement with the theoretical values and resolve the apparent
discrepancy between theory and experiment.
\end{abstract}


\maketitle

The fine structure intervals in helium have attracted a great deal of
interest in recent years because of the possibility of using a
comparison between theory \cite{drake_02,pach} and experiment
\cite{hessels,shiner,inguscio} to better determine the fine structure constant
$\alpha = e^2/\hbar c$. For example, for the $1s2p\;^3{\rm P}_J$ state
of helium ($J=0,1,2$), a measurement of the large interval $\nu_{01}
\simeq 29\,617$ MHz to an accuracy of $\pm$1 kHz is sufficient to
determine $\alpha$ with an accuracy of $\pm$16 ppb (parts per billion).
 Unfortunately, a clear interpretation of the experiment by George {\it
et al.} \cite{hessels} at this level of accuracy is clouded by a rather
large 19.4 kHz disagreement between theory and experiment for the
smaller interval $\nu_{12} \simeq 2291$ MHz \cite{drake_02}.

The situation is further complicated by an apparent discrepancy of
about 250~kHz between theory and a measurement by Yang {\it et al.}
\cite{yang} for the fine structure splittings of the
$1s3p\;^3{\rm P}_J$ state.  The discrepancy is particularly troubling
since all the higher order corrections, which in principle might be
responsible for the discrepancy if they were incorrect, decrease
roughly in proportion to $1/n^3$, where $n$ is the principal quantum
number. Consequently, they are smaller by a factor of $(2/3)^3 \simeq
0.3$ than in the $1s2p\;^3{\rm P}_J$ state.  The purpose of this paper
is to present the results of a more accurate measurement of the fine
structure splittings for the $1s3p\;^3{\rm P}$ state and to compare
them with the theoretical values.

To a first approximation, the various theoretical contributions to the
fine structure splittings come from the spin-orbit (so),
spin-other-orbit (soo) and spin-spin (ss) terms of order $\alpha^2$
Ryd.\ in the Breit-Pauli interaction.  However, comparisons with
experiment at the level of a few kHz also require the calculation of
higher order corrections of order $\alpha^3$, $\alpha^4$,
$\alpha^5\ln\alpha$, and $\alpha^5$.  The complete expression for the
spin-dependent energy shift up to this order is then of the form
\cite{drake_02}
\begin{eqnarray}
\label{deltaE_J}
\Delta E_J &=& \alpha^2\left\langle B_{2,0} \right\rangle + \alpha^3
\left\langle B_{3,0}\right\rangle + \alpha^4 \left\langle B_{2,0}
G'B_{2,0}\right\rangle\nonumber\\[6pt]
&&\mbox{} +\alpha^4\left \langle B_{4,0}\right\rangle+\alpha^5\ln(Z
\alpha)^{-2}\left\langle B_{5,1}^{\rm so}\right\rangle\nonumber\\[6pt]
&&\mbox{}  +2\alpha^5\ln(Z\alpha)^{-2}\left\langle B_{2,0}G'A_{3,1}
\right\rangle\nonumber\\[6pt]
&&\mbox{} 	+\alpha^5\ln\alpha\left\langle B_{5,1}^{\rm soo}  + B_{5,1}^{
\rm ss} \right\rangle  +\alpha^5\left\langle B_{5,0}\right\rangle
\nonumber\\[6pt]
&&\mbox{}  +2\alpha^5\left\langle B_{2,0}G'A_{3,0}\right\rangle + O(
\alpha^6)
\end{eqnarray}
where $A_{m,n}$ and $B_{m,n}$ stand for combinations of relativistic
and QED operators multiplying terms proportional to $\alpha^m(\ln\alpha)
^n$ and $G'$ denotes the reduced Green's function for second-order
contributions.  The leading term $B_{2,0}$ is the standard Breit-Pauli
interaction. $B_{3,0}$ comes from the anomalous magnetic moment and
$B_{4,0}$ contains the sum of 15 Douglas-Kroll operators
\cite{douglas-kroll}.  The terms $B_{5,1}$ and $B_{5,0}$ are quantum
electrodynamic corrections first derived by Zhang \cite{zhang}. The
$B_{5,1}$ part has been independently verified by Pachucki and
Sapirstein \cite{pach}. However, the term $B_{5,0}$ presents
significant theoretical challenges, and it has only partially received
independent verification.  Fortunately, this term is small enough to be
neglected for purposes this work. In addition, there are finite mass
corrections to each of the above terms involving an additional factor
of $\mu/M$ or $(\mu/M)^2$ which must be taken into account, where
$\mu/M$ is the ratio of the reduced electron mass to the nuclear mass.

As discussed previously \cite{drake_02}, the principal computational
step is the calculation of the matrix elements of the various operators
in Eq.\ (\ref{deltaE_J}) with respect to high precision variational wave
functions, and the evaluation of the second-order terms.  The results
for the terms up to and including order $\alpha^5\ln\alpha$, and the
finite mass corrections up to order $\alpha^3\mu/M$, are displayed in
Table~ \ref{tab:FSvalues}. The omitted terms of order $\alpha^5$ and
$\alpha^4\mu/M$ are known in the case of the $1s2p\;^3{\rm P}$ state of
helium to give a net contribution of less than $\pm$20 kHz.  With the
$1/n^3$ scaling, this would reduce to $\pm$6 kHz for the $1s3p\;^3{\rm
P}$ state, which is the value we take as the uncertainty due to the
omitted higher-order terms. This level of accuracy is more than
sufficient for comparison with the accuracy of the measurements
presented here. However, a full calculation of these terms would be
necessary if the experimental accuracy were improved to $\pm$6 kHz, and
a measurement at the $\pm$1 kHz would provide an important test of
theory and resolution of the 19.4~kHz discrepancy in $\nu_{12}$ for the
$1s2p\;^3{\rm P}$ state.
\begin{table}
\caption{Summary of contributions to the fine structure intervals of
the $1s3p\;^3{\rm P}_{J}$ levels in helium.  Units are MHz.}
\label{tab:FSvalues}
\begin{tabular}{l r@{}l r@{}l}
\hline
\multicolumn{1}{c}{Term}&
\multicolumn{2}{c}{$\nu_{01}$}&
\multicolumn{2}{c}{$\nu_{12}$}\\
\hline
$\alpha^2$$\langle B_{2,0}\rangle$ &
8099&.8771   &  666&.1254 \\
$\alpha^2(\mu/M)$$\delta_M\langle B_{2,0}\rangle$  &
--0&.7754   &    0&.8131 \\
$\alpha^2(\mu/M)^2$$\delta_M^{(2)}\langle B_{2,0}\rangle$&
0&.0000   &  --0&.0001 \\
$\alpha^3$$\langle B_{3,0}\rangle$ &
14&.8273   &  --6&.4273 \\
$\alpha^3(\mu/M)$$\delta_M\langle B_{3,0}\rangle$ &
--0&.0020   &    0&.0014 \\
$\alpha^4$$\langle B_{4,0}\rangle$ &
--0&.9654   &    0&.5360 \\
$\alpha^4$$\langle B_{2,0}G'B_{2,0}\rangle$ &
0&.7701(4)&  --2&.2369(5)\\
$\alpha^5\ln(Z\alpha)^{-2}$$\langle B_{5,1}^{\rm so}\rangle$ &
0&.0105   &    0&.0210 \\
$\alpha^5\ln\alpha$$\langle B_{5,1}^{\rm soo}\rangle$ &
--0&.0036   &  --0&.0072 \\
$\alpha^5\ln\alpha$$\langle B_{5,1}^{\rm ss}\rangle$  &
0&.0063   &  --0&.0025 \\
$\alpha^5\ln(Z\alpha)^{-2}$$2\langle B_{2,0}G'A_{3,1}\rangle$&
--0&.0151   &  --0&.0217 \\
$|$higher order terms$|$&                                  $\leq
0$&.0060   & $\leq 0$&.0060\\
Total                                                        &
8113&.7297   &  658&.8012 \\
\hline
\end{tabular}
\end{table}

The new experimental determination of the fine structure intervals
reported here is based on laser induced fluorescence detection in an
atomic beam. Metastable helium atoms are produced in a RF driven
discharge source that is cooled to liquid nitrogen (LN$_2$)
temperature. Two-dimensional transverse cooling on the $1s2s\;^3{\rm S}
_{1} \rightarrow 1s2p\;^3{\rm P}_{2}$ transition at 1083~nm is used to
reduce the divergence of the atomic beam of metastable helium and
increase its forward intensity by a factor of 10. After passing a
collimator and a flight path of 180 cm the atomic beam is finally
overlapped perpendicularly with two antiparallel laser beams at a
wavelength of 389~nm to excite the $1s2s\;^3{\rm S}_{1} \rightarrow
1s3p\;^3{\rm P}_{J}$ ($J=0,1,2$) transitions. A lens images the laser
induced fluorescence onto a photo-multiplier tube (PMT). The overall
efficiency for the photon detection is about 0.1\%. The interaction
region is enclosed by a magnetic shield with an attenuation factor of
800 to minimize possible effects from Zeeman shifts.

The frequency stability and control of the 389~nm laser light is of key
importance for the accurate determination of the fine structure
splitting. The 389~nm light is generated through frequency doubling of
the amplified output of an external-cavity diode laser (DL1) operating
at 778~nm. The frequency of DL1 is locked to a high finesse Fabry-Perot
interferometer (FPI). A frequency tunable acousto-optic modulator (AOM)
is placed between DL1 and the FPI and works as a frequency shifter to
allow for scanning the DL1 frequency relative to a selected FPI mode.
To stabilize the absolute frequency positions of the FPI modes, the FPI
is locked to a second diode laser at 778~nm (DL2), which is itself
referenced to a saturated absorption signal from an iodine cell. The
iodine spectrometer provides a long-term frequency stability of better
than 15~kHz in one minute, which translates to a stability of better
than 30~kHz in one minute for the 389~nm light. Partial beams from DL1
and DL2 are overlapped onto a fast photodiode and the resulting beat
frequency is amplified and continuously monitored by a microwave
frequency counter that is referenced to a rubidium disciplined crystal
oscillator with a relative frequency uncertainty of less than 1 ppb.

The blue laser beam is spatially filtered and expanded to a diameter of
about 1~cm before being sent through the interaction region. A
retroreflector on the opposite side of the vacuum chamber ensures that
the counterpropagating laser beam is in exact anticollinear geometry.
This arrangement largely cancels possible Doppler shifts that would
result from systematic laser beam steering when switching between the
different transitions. Additionally, the laser beam is carefully
aligned to be exactly perpendicular to the atomic beam by matching the
resonance center position obtained with two laser beams to the one
obtained when the return beam is blocked. During the experiment, the
alignment is constantly checked by monitoring the reverse transmission
of the retroreflected beam through the spatial filter pinhole.

\begin{figure}
\includegraphics{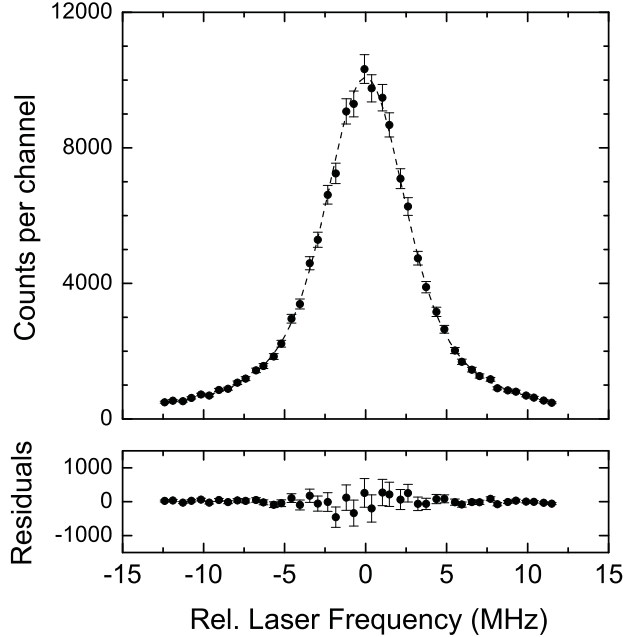}
\caption{\label{fig:resonances} Typical resonance profile for the
$1s2s\;^3{\rm S}_{1}\rightarrow1s3p\;^3{\rm P}_{2}$ transition taken at
a laser power of 50~$\mu$W. The dashed line is a least-square fit of a
Voigt profile to the data. The fit uncertainty on the line center
is 23~kHz. The respective residuals are shown in the lower plot.}
\end{figure}
To obtain a resonance curve, the blue laser frequency is scanned to
cover a range of about $\pm$15~MHz around the fluorescence maximum by
discrete changes of the oscillator frequency that drives the AOM. At
each frequency step, the beat frequency and the PMT count rate are
recorded as data. Fig.~\ref{fig:resonances} shows a typical example of
the resonant curve for the $1s2s\;^3{\rm S}_{1} \rightarrow
1s3p\;^3{\rm P}_{J}$ transition. The PMT count is plotted as a function
of the laser frequency relative to the line center. The errors on the
photon counts are not purely statistical, but also include a 3--5$\%$
contribution from the power fluctuation of the blue laser light during
the counting gate of 0.5~s. Nonlinear least-square fits to the data
using a Voigt profile as a fit function yield the respective center
frequencies. The Voigt profiles fit all curves well with reduced $\chi
^2$ of around one. The residuals plotted in the lower part of
Fig.~\ref{fig:resonances} show mainly statistical scattering around
zero. The values for the fine structure splittings are obtained from
the differences in the center frequencies.

\begin{figure}
\includegraphics{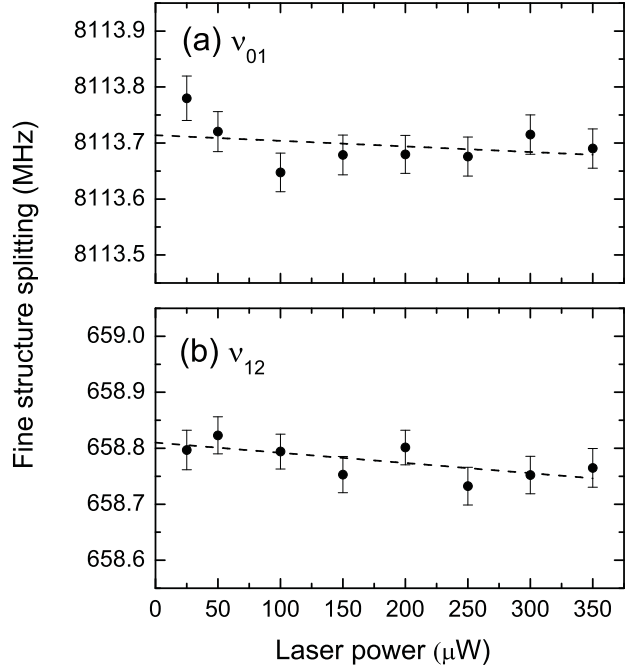}
\caption{\label{fig:powerdependence}Experimental fine-structure
splittings for (a) $\nu_{01} (J=0\rightarrow 1)$ and (b) $\nu_{12} (J=
1\rightarrow 2)$ as a function of laser power; the dashed lines are
linear fits to the data to extract the zero-power values. A laser power
of 350~$\mu$W corresponds to an intensity of $\sim 0.45$~mW/cm$^2$ or
$\sim 13\%$ of the saturation intensity of 3.4~mW/cm$^2$.}
\end{figure}
Momentum transfer from the laser light to the atoms changes their
transverse velocity distribution. Since this effect depends on the
laser frequency, any power imbalance in the two laser beams may result
in small asymmetries of the resonance profile that can lead to
systematic shifts of the fine structure splitting. To minimize this
effect, the laser intensity is chosen to be so low that only a few
photons are scattered by each atom while transversing the laser beam.
Additionally, the fine structure splitting was measured as a function
of the laser power over a range from 25 to 350~$\mu$W or about $0.9\%$
to $13\%$ of the saturation intensity. The corresponding results are
shown in Fig.~\ref{fig:powerdependence}. Each data point is the mean of
at least four independent measurements. The error bars are given by the
standard error of the independent measurements and corresponds well to
the error of the individual peak fitting plus the statistical
fluctuation of the reference laser frequency of about 20~kHz during
one measurement. Overall, the results indeed exhibit a small but
noticeable power dependence as indicated by the respective slope of the
linear regression to the data. The experimental values for the fine
structure intervals are therefore obtained by extrapolating to zero
laser power.

The residual magnetic field within the interaction region was measured
to be less than 3~mGauss. To get an upper estimate for a possible
systematic effect caused by this field strength, the magnetic shield
was removed and a magnetic field was deliberately applied. An asymmetry
of the Zeeman levels of up to $20\%$ at 8~Gauss was observed. From this
measurement and a Zeeman shift of 1.4~MHz/Gauss for the
389~nm transition, it can be inferred that the residual shift on the
line center at 3~mGauss is less than 0.8~kHz. Systematic effects from
light shift and pressure shift are also well below the 1~kHz level. The
only remaining effect of importance is systematic laser beam steering.
The relative alignment of the counterpropagating laser beams provided by
the retroreflecting configuration has a stability of better than
$2.5\times 10^{-3}$~mrad. The maximum Doppler shift under these
conditions is 5~kHz assuming perpendicular geometry and a helium beam
at LN$_2$ temperature.

\begin{figure}
\includegraphics{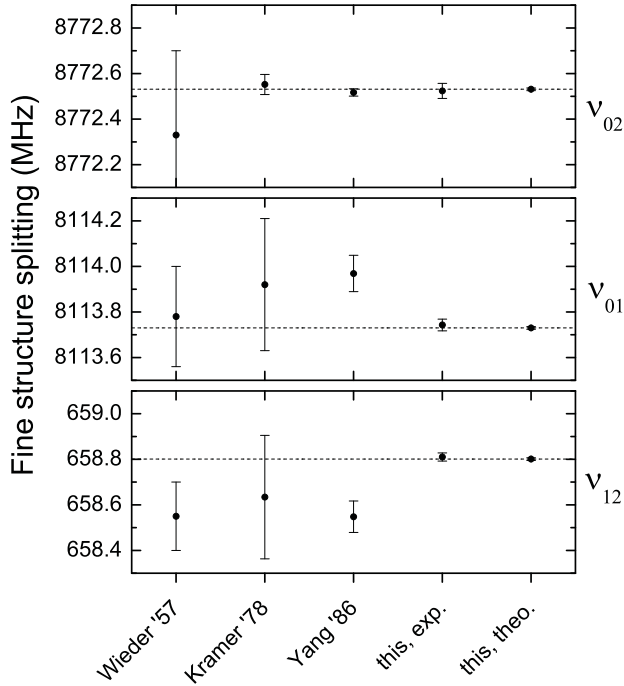}
\caption{\label{fig:compare}Comparison of results for the three fine
structure intervals of the $1s3p\;^3{\rm P}_{J}$ levels from Wieder
{\it et al.} \cite{lamb}, Kramer {\it et al.} \cite{pipkin} and Yang
{\it et al.} \cite{yang}, as well as the experimental and theoretical
values from this work. The theoretical values are also indicated by
the dashed lines.}
\end{figure}

\begin{table}
\caption{Comparison between theoretical and experimental values for the
fine structure intervals of the $1s3p\;^3{\rm P}_{J}$ levels in helium.
Units are MHz.}
\label{tab:CompareFSvalues}
\begin{tabular}{l r@{}l r@{}l r@{}l}
\hline
\multicolumn{1}{c}{Reference}&
\multicolumn{2}{c}{$\nu_{01}$}&
\multicolumn{2}{c}{$\nu_{12}$}&
\multicolumn{2}{c}{$\nu_{02}$}\\
\hline
Wieder \cite{lamb}, exp.  & 8113&.78(22)   &  658&.55(15)   &
8772&.33(37)\\
Kramer \cite{pipkin}, exp.& 8113&.92(29)   &  658&.63(27)   &
8772&.552(40)\\
Yang \cite{yang}, exp.    & 8113&.969(80)  &  658&.548(69)  &
8772&.517(16)\\
this work, exp.           & 8113&.714(28)  &  658&.810(18)  &
8772&.524(33)\\
this work, theory     & 8113&.730(6)   &  658&.801(6)   &
8772&.531(6)\\
\hline
\end{tabular}
\end{table}

The final experimental values for the two independent fine structure
intervals $\nu_{01} (J=0\rightarrow 1$) and $\nu_{12} (J=1\rightarrow
2)$ with their respective combined statistical and systematic
uncertentainties are listed in Table~ \ref{tab:CompareFSvalues}. The
results of previously published measurements on all three fine
structure intervals of the $1s3p\;^3{\rm P}_{J}$ levels as well as the
theoretical and experimental result obtained in this work are compared
in Fig.~\ref{fig:compare}. Our experimental results on the fine
structure splittings agree very well with the theoretical calculations.
While there is a clear disagreement with the results for the $J=0
\rightarrow 1$ and $J=1 \rightarrow 2$ intervals obtained in
\cite{yang}, the value for $J=0 \rightarrow 2$ of both experiments
agree very well. The previous work was based on the measurement of the
magnetic field strength for the crossing point between certain Zeeman
sublevels. The extraction of the fine structure splitting values at
zero field requires an accurate theoretical analysis of the Zeeman
shift \cite{yan_94}. The level crossing of the $J=0$ and $J=1$ levels
occurs at a considerably higher magnetic field than the $J=0$ and $J=2$
crossing. Therefore, the $J=0 \rightarrow 1$ measurement is more
sensitive to systematic uncertainties. Since our new experimental
approach directly determines the fine structure intervals at zero
field, it avoids these complications.

In conclusion, the results from a new experiment measuring the fine
structure intervals of the $1s3p\;^3{\rm P}_{J}$ levels in atomic
helium resolve an apparent discrepancy between theory and the previous
most precise experimental determination. This adds confidence to the
atomic theory calculations of fine structure intervals in atomic helium
that serve as the basis for an atomic physics determination of the fine
structure constant. For both theory and experiment there is room for
improvements. The current limit for theory is the uncertainty in the
$B_{5,0}$ QED term, which will receive further attention in the
future. For the experiment, the next step would be the introduction of
a more stable reference laser and an improvement of the laser coupling
for a reduction of the systematic effect caused by beam steering. Both
should enable measurements down to uncertainties of $\sim 1$~kHz. At
this level, results from the $1s3p\;^3{\rm P}_{J}$ levels would ideally
complement the efforts in improving the $1s2p\;^3{\rm P}_{J}$ data.

\begin{acknowledgments}
We would like to thank R. J. Holt for helpful discussions.
This work was supported by the U.S. Department of Energy, Office of
Nuclear Physics, under Contract No. W-31-109-ENG-38. G. Drake
acknowledges support by NSERC and by SHARCnet.
\end{acknowledgments}


\begin{thebibliography}{99}
\bibitem{drake_02}
G. W. F. Drake, Can.\ J. Phys.\ {\bf 80}, 1195 (2002).
\bibitem{pach}
K. Pachucki and J. Sapirstein, J. Phys.\ B {\bf 35}, 1783 and 3087
(2002).
\bibitem{hessels}
M. C. George, L. D. Lombardi, and E. A. Hessels, Phys.\ Rev.\ Lett. {\bf
87}, 173002 (2001)
\bibitem{shiner}
J. Castillega, D. Livingston, A. Sanders, and D. Shiner, Phys.\ Rev.\
Lett. {\bf 84}, 4321 (2000).
\bibitem{inguscio}
F. Minardi, G. Bianchini, P. C. Pastor, G. Giusfredi, F. S. Pavone, and
M. Inguscio, Phys.\ Rev.\ Lett. {\bf 82}, 1112 (1999); P.C. Pastor, G.
Giusfredi, P. De Natale {\it et al.}, Phys.\ Rev.\ Lett. {\bf 92},
023001 (2004).
\bibitem{yang}
D.-H. Yang, P. McNicholl, and H. Metcalf, Phys.\ Rev.\ A {\bf 33}, 1725
(1986).  See Refs.\ \cite{yan_94} and \cite{marin} for a correction.
\bibitem{douglas-kroll}
M. Douglas and N.M. Kroll, Ann.\ Phys.\ (N.Y.) {\bf 82}, 89 (1974).
\bibitem{zhang}
T. Zhang, Phys.\ Rev.\ A {\bf 53}, 3896 (1996); T. Zhang and G. W. F.
Drake, Phys.\ Rev.\ A {\bf 54}, 4882 (1996); T. Zhang, Z.-C. Yan and
G. W. F. Drake, Phys.\ Rev.\ Lett. {\bf 77}, 1715 (1996).
\bibitem{lamb}
I. Wieder and W. E. Lamb, Jr., Phys.\ Rev.\ {\bf 107}, 125 (1957)
\bibitem{pipkin}
P. B. Kramer and F. M. Pipkin, Phys.\ Rev.\ A {\bf 18}, 212 (1978).
\bibitem{yan_94}
Z.-C. Yan and G. W. F. Drake, Phys.\ Rev.\ A {\bf 50}, R1980 (1994).
\bibitem{marin}
F. Marin, F. Minardi, F.S. Pavone, M. Inguscio, and G.W.F. Drake, Z.
Phys.\ D {\bf 32}, 285 (1995).
\end{thebibliography}
\end{document}